# "Give me the code" - Log Analysis of First-Year CS Students' Interactions With GPT

Pedro Alves[1] [a] and Bruno Pereira Cipriano[1] [b]

[1]*Lusófona University, Portugal*
{*pedro.alves, bcipriano*}@*ulusofona.pt*

Keywords: large language models, programming, gpt, interaction log analysis

Abstract: The impact of Large Language Models (LLMs) like GPT-3, GPT-4, and Bard in computer science (CS) education is expected to be profound. Students now have the power to generate code solutions for a wide array of programming assignments. For first-year students, this may be particularly problematic since the foundational skills are still in development and an over-reliance on generative AI tools can hinder their ability to grasp essential programming concepts. This paper analyzes the prompts used by 69 freshmen undergraduate students to solve a certain programming problem within a project assignment, without giving them prior prompt training. We also present the rules of the exercise that motivated the prompts, designed to foster critical thinking skills during the interaction. Despite using unsophisticated prompting techniques, our findings suggest that the majority of students successfully leveraged GPT, incorporating the suggested solutions into their projects. Additionally, half of the students demonstrated the ability to exercise judgment in selecting from multiple GPT-generated solutions, showcasing the development of their critical thinking skills in evaluating AI-generated code.

## 1 INTRODUCTION

Large Language Models (LLMs) have been shown to have the capacity to generate computer code from natural language specifications (Xu et al., 2022; Destefanis et al., 2023). Currently, there are multiple available LLM-based tools which display that behaviour. Two examples of such tools are OpenAI's ChatGPT [1] and Google's Bard [2].

The implications of such tools for CS education are obvious: students now have access to tools that can generate code to solve programming assignments, with the capacity to obtain full marks or close to it (Prather et al., 2023; Savelka et al., 2023a; Cipriano and Alves, 2023; Cipriano and Alves, 2024; Finnie-Ansley et al., 2023; Ouh et al., 2023; Savelka et al., 2023c; Savelka et al., 2023b; Reeves et al., 2023).

There has been extensive research into how computer science teachers should respond to LLMs, adapting their teaching methods, assessments, and more (Lau and Guo, 2023; Denny et al., 2023b; Daun and Brings, 2023; Leinonen et al., 2023; Liffiton et al., 2023; Finnie-Ansley et al., 2022; Sridhar et al., 2023). Some resist (fight), contemplating ways to prevent students from using chatbots like ChatGPT, such as blocking access or employing detection tools for AI-generated text with questionable effectiveness (OpenAI, 2023). Others embrace this new paradigm, adapting exercises so that students are encouraged to make the most of LLMs, with presentations/discussions or non-text-based prompts (Denny et al., 2023b), or analyzing the tool's capacity to help students (Hellas et al., 2023).

Regardless of the instructional strategy adopted by teachers, be it one of resistance or acceptance, students will inevitably turn to ChatGPT, perhaps with a certain degree of naivety, lacking knowledge about prompting techniques and an exaggerated faith in its responses. Moreover, most of their interactions will occur without direct teacher oversight. With this study, we want to understand what happens in these cases: whether they will succeed in their endeavors or fail miserably, and how can teachers help in their journey.

We investigate two research questions:

**RQ1**: Can first-year students take advantage of LLMs for code generation without any specific or for-mal training?

---
[a]  https://orcid.org/0000-0003-4054-0792
[b]  https://orcid.org/0000-0002-2017-7511
[1] https://chat.openai.com/
[2] https://bard.google.com/chat

**RQ2**: Are students able to incorporate solutions provided by ChatGPT into their projects?

This work makes the following contributions:

- Proposes an exercise template that teachers may use to foster their students' critical thinking skills when interacting with LLM-based tools;

- Analyzes the interactions of 69 first-year CS students with ChatGPT to produce code using the aforementioned template, but without having received any formal training on the matter. This analysis is based on log files provided by the students documenting their attempts to solve a specific exercise within the context of their course project;

- Presents the results of a post-exercise survey (N=52) to find out the students' perceptions on its usefulness.

## 2 RELATED WORK

Some recent studies evaluated students' interactions with LLMs for code generation. One example, (Denny et al., 2023b) analyzed the interactions between students and ChatGPT in several dimensions such as prompt lengths and number of attempts, following a methodology where instead of asking the students to write the code themselves, they had to write prompts to generate the code for them. In (Prather et al., 2023), researchers studied how first year students use Copilot (a GPT-based code generation code, trained on code publicly available on GitHub) to solve a typical assignment on an introductory programming course, and found that these novice students struggle to understand and use Copilot, and are wary of the tool's implications (i.e. such as not being able to understand the generated code), but are optimistic about integrating the tool into their future development workflow. Another study (Prasad et al., 2023) analyzed the GPT interactions of students in a upperlevel course on applied logic and formal methods, using an IDE plugin. The authors of (Babe et al., 2023a) asked 80 students with a single semester of programming background to write prompts for 48 problems and found that the students' prompts can be ambiguous to LLMs. Finally, (Kazemitabaar et al., 2023) analyzed the interaction logs of learners aged 10 to 17, recruited from programming bootcamps, and verified how the availability of Codex (a GPT-based LLM) impacted their ability to independently learn Python. They found evidences of learner's self-regulation, with some learners actively adding code to test the AI-generated code, as well as over reliance, with some learners submitting the assignment's instructions in a single prompt.

Our study differs from these studies since all of them analysed student's interactions using introductory programming exercises as the basis for the study. In our study, students had to generate code for a specific part of a large project (developed over the course of approximately 3 months), meaning that GPT's generated code would have to be integrated with the remaining of the students' code, thus having an extra layer of complexity.

## 3 EXERCISE TEMPLATE

This section describes a template for programming exercises to be solved using Generative AI tools (e.g., ChatGPT, Bard). The structured approach cultivates critical thinking (Naumova, 2023) and requires no prior training in prompt engineering techniques.

The problem can be stated in any form (textual, diagram, etc..) and the students must interact with an LLM to get a good solution, that they can efficiently apply (e.g., incorporate in their course project). Since LLMs are prone to failing (Babe et al., 2023b; Denny et al., 2023a), students must ask the LLM for two alternative implementations, a well-known prompting technique (Treude, 2023), which must be documented within a log file, to be submitted alongside the exercise solution. This log file must also include a concluding section wherein the chosen implementation must be elucidated, along with the rationale behind the selection. In summary, students must follow these steps:

1. State the problem by asking a question to GPT
2. Analyze the provided solution
3. Ask for an alternative solution, eventually providing further information or constraints
4. Compare both solutions
5. Decide on the best solution

All these interactions should be recorded for further analysis.

To ensure uniformity in format and to clarify the expected nature of interaction, an exemplar log file was provided (shown in Listing 1).

## 4 EXPERIMENTAL CONTEXT

This study was performed in the scope of a Data Structures and Algorithms (DSA) course belonging to

Listing 1: Log format example, supplied to the students

```
Human: Please implement a function in Java that
can identify the largest number in an array of
integers.
--------
ChatGPT: Sure thing! Here's a possible
implementation of a function in Java, which finds
the largest number in an array of integers:

public static int findLargest(int[] numbers) {
    int largest = numbers[0]; // start with the
    first number in the array
    for (int i = 1; i < numbers.length; i++) {
        if (numbers[i] > largest) {
            largest = numbers[i]; // update the
            largest number
        }
    }
    return largest;
}

This function takes an array of integers (...)
--------
Human: Implement an alternative version of this
function that doesn't initialize the "largest"
variable with the first number in the array.
--------
ChatGPT: Certainly! Here's a possible alternative
Java implementation to find the largest number in
an array of integers without initializing the
largest variable with the first number in the
array:

public static int findLargest(int[] numbers) {
    int largest = Integer.MIN_VALUE; // start with
     the smallest possible value for an integer
    for (int i = 0; i < numbers.length; i++) {
        if (numbers[i] > largest) {
            largest = numbers[i]; // update the
            largest number
        }
    }
    return largest;
}

This function begins by setting (...)

-------
GROUP'S CONCLUSION: We chose the first option
because it avoids an extra comparison at the
beginning of the loop (the loop can start at
position 1 instead of position 0).
```

a Computer Engineering degree, during the 2022/23 school year. The course takes place in the second semester of the first year, which means that most students have only been exposed to one semester of programming.

The course's main assignment is the development of a project, typically a command line application that performs queries on a very large data set, provided in the form of multiple CSV files. The queries must be implemented using efficient data structures and algorithms—for example, some queries will require the usage of hash tables to achieve good performance. Its development is split in two parts: in the first part, students mainly need to implement code for reading and parsing the input files, and, in the second part, they have to focus on implementing the different queries. This project is either implemented in groups of 2 students or individually.

The project's topic varies from year to year, as does the input file structure and the required queries. In this particular year, the project was about songs and artists, using data extracted from the Million Song data set[3].

Note that, in this course, the professors and instructors allow and even encourage students to interact with GPT. Also, the course's main project explicitly asks students to use GPT in some of the requirements. However, students were warned that they would not be allowed to use GPT during the project's defense at the end of the semester. The following section describes those requirements.

### 4.1 The ChatGPT assignment

To foster a controlled student experimentation of ChatGPT's capabilities and its integration into the educational framework, the project's first part includes a specific ChatGPT exercise.

Students were instructed to employ ChatGPT as an aid in the reading and parsing a large CSV file containing information about artists. The file's format is available[4]. Some of the challenges related to this file were handling two different formats for single-artist songs and multiple-artist songs, handling multiple artists associated with the same song across distinct lines, dealing with invalid lines, among others. There were also technical restrictions: they couldn't use hash-based data structures (e.g., HashMap) in the first part.

Students validated their code using an automated assessment tool (AAT). They could submit multiple times without penalty and received the execution results of several unit tests.

To guide their interaction with ChatGPT, this exercise followed the template proposed on Section 3 and resulted in a log file detailing all their interactions to achieve a good solution. Incorporating this log file within the project was optional but carried a weight of 0.5 points on the evaluation scale ranging from 0 to 20.

---

[3]http://millionsongdataset.com/pages/getting-dataset/
[4]https://doi.org/10.5281/zenodo.8430808

This strategic integration of ChatGPT served to not only motivate students to delve into its usage under controlled conditions but also to encourage a structured evaluation process.

## 5 METHODOLOGY

In order to better understand how the students interacted with ChatGPT to solve the ChatGPT assignment (see Section 4.1), we manually analyzed their interaction logs, categorizing them in several dimensions. Furthermore, a brief survey was administered subsequent to the assignment to ascertain the sentiments of the students towards the task.

### 5.1 Categorization

Since the students had been instructed to interact with ChatGPT using the exercise template described in Section 3, their logs mostly followed these steps: (1) ask a question; (2) ask for an alternative solution; and (3) provide a conclusion. Therefore, we decided to analyze each step individually: **initial prompt**, **second prompt** and **conclusion**. We added a fourth dimension called **problem**, to capture the diversity of the problems students were asking GPT to help them for.

We then further categorized each dimension into the following categorization variables:

- **Problem** - What was the problem students were asking GPT help for?
  >**Abstraction level** (high or low) - Some students asked very concrete and direct questions such as '*How to remove spaces and quotes from a string in Java?*' while others asked more abstract questions such as '*Implement a function in Java that read a .txt file with the following format (...)*'
  >**Nature** (generic or domain-specific) - If the stated problem was or wasn't specific to the project. For example '*How to remove spaces and quotes from a string in Java?*' is a generic problem, since it can be applied to a variety of problems and not specifically to the 'artists file parsing' problem

- **Initial prompt**
  >**Language** - Since our students are native Portuguese speakers, the majority of the interactions were done in Portuguese, although a small percentage of students used English

  >**Type** - This defines the goal of the prompt
  – **Ask for code** - These prompts usually included terms like '*implement a function...*' or '*give me the code...*' which inspired the title of this paper
  – **Explain how** - Ask GPT to explain how they could solve a certain problem, without explicitly asking for code
  – **Help with error** - Students provided an error they were struggling with (e.g. a compilation error which they couldn't understand)
  – **Analyze code** - Students provided code from their project and asked GPT for errors or possible improvements in their code.

  >**Provided context?** - Some students copied or adapted part of the project statement explaining the rules behind the artists file, including them into the initial prompt.

  >**Provided restrictions** - Here we analyzed if the students provided ChatGPT with technical restrictions. In particular, we looked for the words 'java' and 'hashset/hashmap' in the prompt. The first was the language the project had to be developed in and the second was an explicit prohibition in the project statement (since at this point, they hadn't yet learnt these data structures).

  >**Gave examples?** - For prompts of the type 'ask for code', we analyzed if the students provided any examples to guide the model. We considered examples of input (e.g., lines in the artists file) and/or output (e.g., given these parameter values, the function should return this).

  >**Function signature** - For prompts of the type 'ask for code', we analyzed if the students provided the function signature beforehand (i.e., the name of the function along with its parameters and return type).

- **Second prompt**
  >**Type**
  – **Inexistent** - Although the exercise template directed students to ask for alternatives, some students provided only the initial prompt
  – **Give me an alternative** - We analyzed if the students asked for an alternative solution. This was one of the goals of this exercise.
  – **Clarify initial prompt** - We analyzed if the second prompt was just a clarification of the first prompt. This could be an attempt to generalize the previous solution or provide more information.
  – **Ask different question** - Some students didn't comply with the exercise statement and used the second prompt to ask for a different question

>**Guided alternative?** - Here we wanted to find out if, only for the 'give me an alternative' cases, the students provided some guidance (e.g., '*Implement again without using replaceAll()*') when asking for an alternative solution

- **Conclusion**
  >**Type**
  - **Decided for one solution** - We analyzed if the students achieved the main goal of the exercise, by choosing one of the alternative solutions given by ChatGPT
  - **Useful but didn't use the solution** - Some students didn't use any of the solution but have written they still found the interaction to be useful (e.g., '*We didn't use any of the solutions. However we used some of the suggested instructions such as* `replaceAll()`')
  - **Not useful** - Some students didn't find any of the solutions useful
  - **Inexistent/invalid conclusion** - There are the cases where the students didn't write any conclusion or the conclusion seems to have no connection with the given prompts.

  >**Decision criteria** - This metric was applied only to the conclusions of the type 'Decided for one solution'. All the decisions fell into one of these categories: simpler/more compact, more flexible/generic, and easier to understand.
  >**Used GPT solution** - Here we wanted to find out if the students really used the GPT solution (complete ou partially) in their projects. We verified this by manually analyzing the project's code for each group, searching for code that is similar to one of the GPT solutions.

### 5.1.1 Classification example

As an example, the interaction shown in Listing 2 was classified like this:

**Problem**: low abstraction level; general.

**First prompt**: 'ask for code' type; no context given; provided a restriction ("must be in Java"); provided an example but didn't provide the requested function signature

**Second prompt**: inexistent

**Conclusion**: useful but didn't directly use the provided solution; we confirmed that their project actually used one of the suggestions made by ChatGPT regarding the use of replaceAll().

## 5.2 Survey

To complement the log analysis results, in particular regarding the usefulness of the ChatGPT assignment

Listing 2: Partial example of log file supplied by the students (group 22), translated from Portuguese.

```
Human: How to remove the white spaces between the
symbols [, ', and ] in Java?
1434545625 @ [ 'The National' ]
--------
ChatGPT: To remove the white spaces between the
symbols [, ', and ] in a string in Java, you can
use regular expressions and string replacement.

Here's an example of code that uses regular
expressions to remove these white spaces:
(...)
Explanation:
The first replaceAll replaces the sequence (...)
-------
Group's Conclusion: We didn't use what ChatGPT
provided, but we ended up using the function that
it suggested: replaceAll()
```

(see Section 4.1), we conducted an anonymous questionnaire a few weeks after the completion of the assignment. Of the 154 enrolled students, 52 responded to the questionnaire, corresponding to a 33.77% participation rate.

The questionnaire was composed of the following 4 questions:

**Q1** - In part 1 of the project, you were asked to interact with GPT. Did you? [Yes/No];

**Q2** - If you answered "No" to the previous question, what is the reason for not interacting with GPT? [Open-ended];

**Q3** - If not for the exercise, would you still have used GPT? [Yes/No];

**Q4** - How useful do you think this exercise was (asking ChatGPT for help in processing the artists' file)? [Scale:1-5].

## 6 RESULTS

65 groups (corresponding to 122 students) submitted a project that passed at least half of the assessment tests. From those, 37 groups (corresponding to 69 students), submitted a log file of the respective interaction with ChatGPT.

We manually analyzed each of those files both quantitatively and qualitatively. A public repository of these files is available[5].

---

[5]https://doi.org/10.5281/zenodo.8430808

Table 1: Log analysis results: problem

| problem | abstraction level | high | 52.8% |
|---|---|---|---|
| | | low | 47.2% |
| | nature | generic | 35.1% |
| | | domain | 64.9% |

Table 2: Log analysis results: initial prompt.

| initial prompt | language | portuguese | 91.7% |
|---|---|---|---|
| | | english | 8.3% |
| | type | ask for code | 69.4% |
| | | explain how | 19.4% |
| | | help error | 8.3% |
| | | analyze code | 2.9% |
| | **provided context?** | yes, copied | 9.1% |
| | | yes, adapted | 9.1% |
| | | no | 81.8% |
| | **provided restrictions** | must be in java | 81.3% |
| | | must not use hashset | 9.1% |
| | | none | 21.2% |
| | **gave examples?** | no | 71.9% |
| | | just one | 9.4% |
| | | several | 18.7% |
| | **function signature?** | no | 92.9% |
| | | yes | 7.1% |

## 6.1 Quantitative

We classified all the ChatGPT interaction logs following the criteria outlined in Section 5, for the 4 steps involved: problem, first prompt, second prompt and conclusion.

As illustrated by Table 1, the abstraction level of the **problem** presented to ChatGPT was evenly distributed between low and high, with a slight inclination toward the high level (52.8%). The majority of students posed domain-specific questions (64.9%) rather than generic ones. These results were expected, given the context of using ChatGPT within a specific project.

Regarding the **initial prompt** (see Table 2), most of the students used their native language (91.7%) and asked for code (72.2%). Still, a small minority preferred to ask ChatGPT to explain how they could solve the problem (19.4%), and an even smaller fraction asked for help identifying errors (8.3%) or analyzing code (5.6%). Most students didn't provide any context (81.8%) but provided restrictions, mainly the restriction that it had to be in Java (81.3%). Also, the majority of the students didn't provide examples (71.9%) and almost no students provided the signature of the pretended function (92.9%). Notice that these last 2 criteria were only applied to interactions where the students asked for code (which most did).

In reference to the **second prompt** (see Table 3), a significant number of students didn't provide it (27.8%) which was surprising since the students were explicitly instructed to do so. Still, the majority asked for an alternative solution on the second prompt (61.1%) as instructed. Interestingly enough, a few students opted to use the second prompt to clarify what they wanted in the initial prompt (11.1%) and even fewer students just asked a different question (5.6%). For cases in which students requested an alternative solution, it is noteworthy that the majority did not provide any specific guidance for their request, merely asking for a generic alternative (71.4%). Only in 28.6% of the cases did the students provide some guidance by adding more information.

Finally, with respect to the **conclusion** (see Table 4), almost half of the students accepted one of the solutions provided by ChatGPT (47.2%) and for those that haven't, a significant portion still found the interaction to be useful (25%), with only a small minority not getting any value from the interaction (11.1%). For the cases where the students accepted one of the solutions, the findings unveil a balanced distribution of the acceptance criteria. A substantial 35.7% of respondents valued simplicity and compactness as guiding factors. Similarly, an equal 35.7% sought flexibility and generality. Finally, 28.6% prioritized ease of understanding. Regardless of what the students wrote in the conclusion, we found that 72.2% really used (fully or partially) one of the solutions provided by ChatGPT.

**General insights** The exercise template encouraged a rigid methodology for using ChatGPT: students were instructed to use it as assistance in reading and parsing the artist file, obtain two alternative implementations, and apply critical thinking to choose one while justifying their decision. However, how students utilized it varied considerably. Some chose to focus on a highly specific problem (low abstraction level), as seen in Listing 2, while others posed more generic problems, with no clear trend towards either option.

The initial prompting methods employed by students also displayed a significant degree of diversity. This was expected as they had no formal training, and the project statement did not provide any guidance in this regard. Nonetheless, the majority of prompts were aimed at obtaining code, although a substantial portion of students also requested explanations on how to proceed. Due to their limited training, these prompts tended to be unsophisticated, lacking context, examples, or the pretended function signature. Even in terms of restrictions, while most students remembered to specify Java, few remembered to indicate that the use of hashsets was not allowed. However, this didn't stop most students from getting useful results (RQ1): specifically, 47.2% of participants acknowledged acceptance of one of the provided solutions, with an additional 25% expressing utility despite abstaining from utilization.

Table 3: Log analysis results: second prompt.

| | | | |
|---|---|---|---|
| second prompt | **type** | inexistent | 27.8% |
| | | give me an alternative | 61.1% |
| | | clarify initial prompt | 11.1% |
| | | ask different question | 5.6% |
| | **guided alternative?** | yes (more information, restrictions, ...) | 28.6% |
| | | no, just "give me an alternative" | 71.4% |

Table 4: Log analysis results: conclusion.

| | | | |
|---|---|---|---|
| conclusion | **type** | decided for one solution | 47.2% |
| | | useful but didn't use the solution | 25% |
| | | was not useful | 11.1% |
| | | inexistent/invalid conclusion | 13.9% |
| | **decision criteria** | simpler/more compact | 35.7% |
| | | more flexible/generic | 35.7% |
| | | easier to understand | 28.6% |
| | **used gpt solution** | yes | 44.4% |
| | | partially | 27.8% |
| | | no | 27.8% |

A considerable number of students (38.9%) did not adhere to the project's instructions and failed to request an alternative implementation. The reasons for this remain unclear, but we suspect it may be due to (1) students not carefully reading the project statement, and (2) being an uncommon way to use ChatGPT.

For the students who did request an alternative solution, it is interesting to note that a substantial portion (28.6%) attempted to guide the solution by specifying constraints, despite receiving no indication to do so.

In the conclusions, it was expected that students would employ their critical thinking skills to choose one of the implementations, which occurred in half of the cases. Even in instances where this did not occur, a substantial portion of students found the interaction to be valuable. A confirmation of LLM's effectiveness of in assisting students is evident, with only 27,8% not being able to incorporate solutions provided by ChatGPT into their projects (RQ2).

Also, half of the students managed to obtain two alternative solutions, with no clear winner among the decision criteria.

### 6.2 Qualitative

During the classification process, we found some cases worth discussing individually.

#### 6.2.1 The Database prompt

Group 36 provided the following initial prompt:
*I must create a Java database that receives three*

*files: songs, artists, and artist details. I need to associate songs with artists because there must be something that relates them, as one artist can have multiple songs, and multiple songs can belong to other artists, meaning it is a many-to-many (N:N) relationship. How do I program this?*

This prompt mixes concepts from Programming and Databases, it seems the student wanted to create a database but it ends with 'How do I program this?'. ChatGPT answered with a program that interacts with a relational database, which was not the goal of the project. What we found interesting in this prompting is its high abstraction level, for sure the highest on all of the prompts used by the students. However, this led to a solution that while correct, could not be used in the project. Notably, this student was the only one who had taken the Databases course, which is taught in the second year, while most of these students were first-year students.

### 6.2.2 Incomplete first prompts

Some groups used the second prompt to complement the first prompt, since the results were not satisfactory. This was accounted in the 'clarify first prompt' item, in Table 3.

For example, Group 9 issued the following prompt: *I want to read a line that has this format "['Name1', 'Name2']." How can I check in Java if the given String follows this format?*

Since ChatGPT's response was specific for lines with two names, the group issued a second prompt asking for a more general solution: *What if I have to handle more than two names, what do I do?*

This second prompt could be prevented if the students had asked for a general solution in the first prompt, or provided several examples instead of just one. This confirms that some students cannot get the best results out of ChatGPT without formal training [RQ1].

### 6.2.3 Arm wrestling with ChatGPT

Group 32 entered into a kind of "arm wrestling" with ChatGPT since they were not getting the answer they needed. They asked ChatGPT to explain what could be improved in a function they developed and that was probably not working as expected. ChatGPT suggested that they could remove some code duplication. However, this wasn't the cause of the error; in fact, there was no code duplication at all. After several prompts from the students, ChatGPT kept insisting the there was code duplication (possible "hallucination") and the students kept rebutting, without success. Here is one of the prompts:

*so, the issue is on a function that I showed you after you told me there was code duplication? Are you sure about this? what about the first piece of code i showed you? (...)*

Our analysis of the students' function revealed no code duplication issues, indicating that ChatGPT's answers were misleading. Interestingly, the students couldn't find the code duplication nor conceive that ChatGPT was wrong, so they kept pushing for more information.

### 6.2.4 Why am I failing the automated assessment tests?

Group 4's motivation for asking for help was the fact that they weren't passing an automated assessment test and they were not understanding why. That particular assessment test exercised edge cases in the requirements which were not taken proper care in their implementation such as duplicate ids and missing information. Notice that the source code of the automated tests wasn't available to the students.

After describing the artists file format, the students asked ChatGPT for examples of input files that would cover all possible scenarios. Although ChatGPT provided some examples, they were not exhaustive so it didn't help them pass the tests. The students kept pushing ChatGPT for more scenarios with no success. At a certain point, they switched their strategy, explaining to ChatGPT how they had implemented the function and asking for what could be missing. Still, ChatGPT wasn't able to help them.

We found interesting that this group opted to implement the function by themselves, abstaining from assistance provided by ChatGPT. Subsequently, upon encountering difficulties in passing the automated assessment tests, they reached out to ChatGPT, albeit without success.

## 6.3 Survey

This section presents the results of the survey conducted after the students completed the exercise (N=52).

84.6% (44) of the students replied that they interacted with GPT during the project (Q1). The students that did not interact (15.4%, 8) provided two main reasons for not doing so (Q2): 6 students replied "I didn't feel the need to", while 2 students replied "I didn't know how to".

Question (Q3) tried to understand if the students would have used GPT even if there was no requirement for doing so. Most students, 69.2% (36) indicated that they would have used GPT anyway, but 30.8% (16) indicated that they would not have used

GPT unless asked to do so. This surprised us, as we expected more students to indicate they would have used GPT anyway.

The fourth question (Q4) aimed to assess the perceived utility of the ChatGPT assignment within the framework of students' projects. Among the respondents, 5.8% (3) selected Option 1 ('Useless'), while 21.2% (11) opted for Option 2 ('Slightly useful'). Options 3 and 4 were each chosen by 30.8% (16) of participants. Additionally, Option 5 ('Very useful') was selected by 11.5% (6) of students. These findings indicate a prevalent perception of usefulness among the majority of students, although opinions vary considerably. However, it is noteworthy that a substantial proportion (27%) regarded the exercise as minimally or not useful. We hypothesize that this subgroup may include students who struggled to elicit useful solutions from GPT to apply in their projects.

In conclusion, a vast majority of students (84.6%) will use GPT in an assignment if directed to do so, and most students will use it even if it's not asked of them, but the percentage is lower (69.2%). Most students (73%) found the assignment to be useful or very useful.

# 7 LESSONS LEARNT

Based on this experiment, we leave some recommendations for teachers wanting to incorporate LLMs in their classes.

Integrate **prompt training** into their curriculum, focusing on providing richer information to LLMs, such as the problem context, examples, function signatures, etc. Literature suggests that providing examples tends to improve LLM's results (Brown et al., 2020);

Incorporate designated **LLM-based exercises** into students' projects, encouraging the use of these tools as aids for tackling complex problems. Certain students may experience discomfort in utilizing LLMs, either due to a lack of familiarity with their operation or apprehension regarding their appropriateness for use. In such cases, they may benefit from gentle encouragement or guidance to overcome these barriers.

Use exercise templates that guide students into **approaching LLMs with a critical thinking mindset**. This can be done by asking for multiple solutions and selecting one of them using criteria such as flexibility, compactness or ease of understanding. Evaluating multiple LLM-generated solutions is a skill that we, as well as other researchers (Treude, 2023; Alves and Cipriano, 2023), believe will be of great importance in the future.

# 8 LIMITATIONS

GPT's behaviours are not deterministic. Furthermore, research has shown that they are dynamic, and can vary greatly over time (Chen et al., 2023). As such, it is hard to generalize conclusions, since some students might have had better results than other students, not due to lack of 'prompting skill', but due to the models themselves.

We expect most of the students to have used ChatGPT with GPT model version 3.5, due to it being free. However, some students might have used the GPT-4 model[6]. Since we do not have this information, our results were not controlled for it.

# 9 Conclusions

The proposed exercise template mostly achieved its goal of fostering students' critical thinking when interacting with LLMs, enabling them to generate more refined solutions. Our log analysis shows that most students can effectively use GPT without formal training. With 72.2% of students incorporating ChatGPT's solutions into their projects, we consider the exercise successful. Additionally, our survey (N=52) indicates that 73% of students found the exercise useful, with some stating they wouldn't have used GPT otherwise. However, the limited sophistication in prompting highlights the need for further training, as students often required many prompts or failed to achieve satisfactory results.

More research is needed to devise effective strategies for instructing CS students in the proper utilization of these tools. It is crucial to convey an awareness of their limitations and discourage over-reliance, ultimately better preparing for professional life.

# REFERENCES

Alves, P. and Cipriano, B. P. (2023). The centaur programmer–How Kasparov's Advanced Chess spans over to the software development of the future. *arXiv preprint arXiv:2304.11172*.

Babe, H. M., Nguyen, S., Zi, Y., Guha, A., Feldman, M. Q., and Anderson, C. J. (2023a). Studenteval: A benchmark of student-written prompts for large language models of code. *arXiv preprint arXiv:2306.04556*.

---

[6]GPT-4 was released at our location (Europe) in March, 14, 2023, while our course was running.

Babe, H. M., Nguyen, S., Zi, Y., Guha, A., Feldman, M. Q., and Anderson, C. J. (2023b). StudentEval: A Benchmark of Student-Written Prompts for Large Language Models of Code. arXiv:2306.04556 [cs].

Brown, T., Mann, B., Ryder, N., Subbiah, M., Kaplan, J. D., Dhariwal, P., Neelakantan, A., Shyam, P., Sastry, G., Askell, A., Agarwal, S., Herbert-Voss, A., Krueger, G., Henighan, T., Child, R., Ramesh, A., Ziegler, D., Wu, J., Winter, C., Hesse, C., Chen, M., Sigler, E., Litwin, M., Gray, S., Chess, B., Clark, J., Berner, C., McCandlish, S., Radford, A., Sutskever, I., and Amodei, D. (2020). Language Models are Few-Shot Learners. In *Advances in Neural Information Processing Systems*, volume 33, pages 1877–1901. Curran Associates, Inc.

Chen, L., Zaharia, M., and Zou, J. (2023). How is ChatGPT's behavior changing over time? *arXiv preprint arXiv:2307.09009*.

Cipriano, B. P. and Alves, P. (2023). GPT-3 vs Object Oriented Programming Assignments: An Experience Report. In *Proceedings of the 2023 Conference on Innovation and Technology in Computer Science Education V. 1*, pages 61–67.

Cipriano, B. P. and Alves, P. (2024). LLMs Still Can't Avoid Instanceof: An investigation Into GPT-3.5, GPT-4 and Bard's Capacity to Handle Object-Oriented Programming Assignments. In *Proceedings of the IEEE/ACM 46th International Conference on Software Engineering: Software Engineering Education and Training (ICSE-SEET)*.

Daun, M. and Brings, J. (2023). How ChatGPT Will Change Software Engineering Education. In *Proceedings of the 2023 Conference on Innovation and Technology in Computer Science Education V. 1*, pages 110–116.

Denny, P., Kumar, V., and Giacaman, N. (2023a). Conversing with Copilot: Exploring Prompt Engineering for Solving CS1 Problems Using Natural Language. In *Proceedings of the 54th ACM Technical Symposium on Computer Science Education V. 1*, pages 1136–1142, Toronto ON Canada. ACM. SIGCSE 2023.

Denny, P., Leinonen, J., Prather, J., Luxton-Reilly, A., Amarouche, T., Becker, B. A., and Reeves, B. N. (2023b). Promptly: Using Prompt Problems to Teach Learners How to Effectively Utilize AI Code Generators. arXiv:2307.16364 [cs].

Destefanis, G., Bartolucci, S., and Ortu, M. (2023). A Preliminary Analysis on the Code Generation Capabilities of GPT-3.5 and Bard AI Models for Java Functions. *arXiv preprint arXiv:2305.09402*.

Finnie-Ansley, J., Denny, P., Becker, B. A., Luxton-Reilly, A., and Prather, J. (2022). The robots are coming: Exploring the implications of openai codex on introductory programming. In *Proceedings of the 24th Australasian Computing Education Conference*, pages 10–19.

Finnie-Ansley, J., Denny, P., Luxton-Reilly, A., Santos, E. A., Prather, J., and Becker, B. A. (2023). My ai wants to know if this will be on the exam: Testing openai's codex on cs2 programming exercises. In *Proceedings of the 25th Australasian Computing Education Conference*, pages 97–104.

Hellas, A., Leinonen, J., Sarsa, S., Koutcheme, C., Kujanpää, L., and Sorva, J. (2023). Exploring the Responses of Large Language Models to Beginner Programmers' Help Requests.

Kazemitabaar, M., Hou, X., Henley, A., Ericson, B. J., Weintrop, D., and Grossman, T. (2023). How novices use llm-based code generators to solve cs1 coding tasks in a self-paced learning environment.

Lau, S. and Guo, P. (2023). From "Ban it till we understand it" to "Resistance is futile": How university programming instructors plan to adapt as more students use AI code generation and explanation tools such as ChatGPT and GitHub Copilot.

Leinonen, J., Denny, P., MacNeil, S., Sarsa, S., Bernstein, S., Kim, J., Tran, A., and Hellas, A. (2023). Comparing code explanations created by students and large language models. *arXiv preprint arXiv:2304.03938*.

Liffiton, M., Sheese, B., Savelka, J., and Denny, P. (2023). CodeHelp: Using Large Language Models with Guardrails for Scalable Support in Programming Classes.

Naumova, E. N. (2023). A mistake-find exercise: a teacher's tool to engage with information innovations, ChatGPT, and their analogs. *Journal of Public Health Policy*, 44(2):173–178.

OpenAI (2023). How can educators respond to students presenting ai-generated content as their own? https://help.openai.com/en/articles/8313351-how-can-educators-respond-to-students-presenting-ai-generated [Online; last accessed 03-October-2023].

Ouh, E. L., Gan, B. K. S., Shim, K. J., and Wlodkowski, S. (2023). Chatgpt, can you generate solutions for my coding exercises? an evaluation on its effectiveness in an undergraduate java programming course. *arXiv preprint arXiv:2305.13680*.

Prasad, S., Greenman, B., Nelson, T., and Krishnamurthi, S. (2023). Generating Programs Trivially: Student Use of Large Language Models. In *Proceedings of the ACM Conference on Global Computing Education Vol 1*, pages 126–132, Hyderabad India. ACM.

Prather, J., Reeves, B. N., Denny, P., Becker, B. A., Leinonen, J., Luxton-Reilly, A., Powell, G., Finnie-Ansley, J., and Santos, E. A. (2023). "It's Weird That it Knows What I Want": Usability and Interactions with Copilot for Novice Programmers. *ACM Transactions on Computer-Human Interaction*, 31(1):1–31.

Reeves, B., Sarsa, S., Prather, J., Denny, P., Becker, B. A., Hellas, A., Kimmel, B., Powell, G., and Leinonen, J. (2023). Evaluating the performance of code generation models for solving parsons problems with small prompt variations. In *Proceedings of the 2023 Conference on Innovation and Technology in Computer Science Education V. 1*, pages 299–305.

Savelka, J., Agarwal, A., An, M., Bogart, C., and Sakr, M. (2023a). Thrilled by Your Progress! Large Language Models (GPT-4) No Longer Struggle to Pass Assessments in Higher Education Programming Courses.

Savelka, J., Agarwal, A., Bogart, C., and Sakr, M. (2023b). Large Language Models (GPT) Struggle to Answer Multiple-Choice Questions about Code. *arXiv preprint arXiv:2303.08033*.

Savelka, J., Agarwal, A., Bogart, C., Song, Y., and Sakr, M. (2023c). Can Generative Pre-trained Transformers (GPT) Pass Assessments in Higher Education Programming Courses? In *Proceedings of the 2023 Conference on Innovation and Technology in Computer Science Education V. 1*. ACM.

Sridhar, P., Doyle, A., Agarwal, A., Bogart, C., Savelka, J., and Sakr, M. (2023). Harnessing llms in curricular design: Using gpt-4 to support authoring of learning objectives.

Treude, C. (2023). Navigating Complexity in Software Engineering: A Prototype for Comparing GPT-n Solutions. arXiv:2301.12169 [cs].

Xu, F. F., Alon, U., Neubig, G., and Hellendoorn, V. J. (2022). A Systematic Evaluation of Large Language Models of Code. In *Proceedings of the 6th ACM SIGPLAN International Symposium on Machine Programming*, pages 1–10, San Diego CA USA. ACM.